\documentclass[aps,prl,twocolumn,
showpacs,preprintnumbers,amsmath,amssym,superscriptaddress]{revtex4}
\usepackage{graphicx}
\usepackage{bm}

\begin{document} 

\title{Langevin description of critical phenomena with two symmetric absorbing states}

\author{Omar Al Hammal}
\affiliation{Instituto~de~F\'\i sica~Te\'orica~y~Computacional~Carlos~I,
~Facultad~de~Ciencias,~Universidad~de~Granada,~18071~Granada,~Spain}

\author{Hugues Chat\'e}
\affiliation{CEA -- Service de Physique de l'\'Etat Condens\'e,~CEN~Saclay,~91191
~Gif-sur-Yvette,~France}

\author{Ivan Dornic}
\affiliation{CEA -- Service de Physique de l'\'Etat Condens\'e,~CEN~Saclay,~91191
~Gif-sur-Yvette,~France}

\author{Miguel A. Mu\~noz}
\affiliation{Instituto~de~F\'\i sica~Te\'orica~y~Computacional~Carlos~I,
~Facultad~de~Ciencias,~Universidad~de~Granada,~18071~Granada,~Spain}

\date{\today}

\begin{abstract}
On the basis of general considerations, we propose a Langevin equation
accounting for critical phenomena occurring in the presence of two
symmetric absorbing states. We study its phase diagram by mean-field
arguments and direct numerical integration in physical dimensions.
Our findings fully account for and clarify the intricate picture known
so far from the aggregation of partial results obtained with
microscopic models. We argue that the direct transition from disorder
to one of two absorbing states is best described as a (generalized)
voter critical point and show that it can be split into an Ising and a
directed percolation transitions in dimensions larger than one.
\end{abstract}

\maketitle
The classification of equilibrium phase transitions into universality
classes by just identifying their relevant ingredients
(i.e. symmetries, conservation laws, and dimensionalities) constitutes
one of the most remarkable achievements of modern statistical
mechanics. Ginzburg-Landau-Wilson free-energy functionals, either in
their static \cite{Amit} or dynamic versions (usually written in terms
of Langevin equations, as the Model A of Hohenberg and Halperin
describing the kinetic Ising class  \cite{HH}) provide a compact and
systematic theoretical framework to represent universality classes:
Being continuous (coarse-grained) theories, they are thus susceptible
to analytical studies by using the tools of statistical field theory
and the renormalization group.

Out of equilibrium, the situation is far from being as
satisfactory. In spite of evidence of universality, the relevant
ingredients for classification are often not known, and continuous
descriptions in terms of Langevin equations or dynamical generating
functionals are mostly lacking. For instance, within the prototypical
case of absorbing phase transitions, where the ordered ``absorbing''
states are devoid of fluctuations allowing the return to the
disordered ``active'' ones, the directed percolation (DP) class is
prominent and very robust \cite{RFT,Review_Haye}.
  Loosely defined as the class of all phase
transitions into a single effective absorbing state without extra
symmetries or conservation laws, it is represented by a Langevin
equation which can be renormalized satisfactorily
\cite{RFT,Canet}.  But such a continuous description or
even a controlled renormalisation procedure is lacking for the also
rather well-established class of phase transitions into one of two
symmetric absorbing states, despite some thoughtful attempts
\cite{TC98}.

In this Letter, we propose, on the basis of general
considerations, a Langevin equation accounting for critical
phenomena with two ($Z_2$-)symmetric absorbing states. Pending a
renormalisation group approach, we study the phase diagram of this
equation by mean-field arguments and direct numerical integration and
show that it fully accounts for the rather intricate picture known so
far from the aggregation of partial results obtained with microscopic
models, a situation which we briefly recall now before proceeding with
our findings.

\begin{figure}
\includegraphics[width=3.8cm,clip]{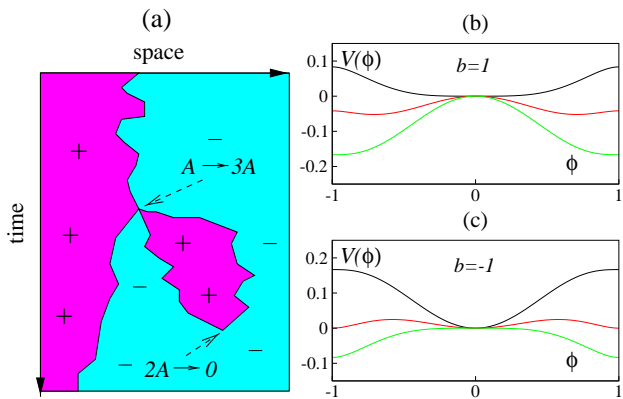}
\includegraphics[width=4.3cm,clip]{./fig1bc.eps}
\caption{(Color online) 
(a) Schematic correspondence between domain and interface 
representation of one-dimensional generalized voter dynamics. 
Here an interface separating a $+$ from a $-$
domain branches ($A\to 3A$, swap of two spins $+-\to-+$), 
creating two new domains; later, two of 
the three interfaces annihilate ($2A\to\emptyset$), suppressing a $+$ domain.
The two symmetric absorbing states impose parity conservation of interfaces.
(b) variation of the deterministic potential with $a$ for $b=1$
(from top to bottom $a=0$, $0.5$ and $1$).
(c) same as (b) but for $b=-1$ and $a=-1$, $-1/3$, $0$.}
\label{fig1}
\end{figure}

The lack of consensus about the characterization of phase transitions
into two symmetric absorbing states is reflected by the different
names given to this class in the literature
\cite{Review_Haye,PC,Hungary,DP2,Review_Odor}. Sometimes simply called
DP2 (marking the existence of two absorbing states), or, more
accurately, ``directed Ising'' (referring to both the Ising
$Z_2$-symmetry and the presence of absorbing states), it is most often
called ``parity-conserving'' because of its usual interface
representation where, in one space dimension,  diffusing particles $A$
undergo the reactions $A\to 3A$, $2A\to\emptyset$.
 (These branching and
annihilating random walks with an even number of offsprings stand for
interfaces between domains of the $+$ and $-$ absorbing states, see
Fig.~\ref{fig1}a.). However, there now exists ample evidence that the
conservation of the parity (of the number of interfaces or particles) is
{\it not} the relevant ingredient \cite{DP2,Review_Haye}. Moreover, this
particle representation just gives rise to trivial phase transitions
(at zero branching rate and with mean-field exponents) in
 higher space dimensions $d \ge 2$.
Rather, as was briefly hinted at in \cite{Haye_DPn}
and suggested in \cite{Kockel1}, 
we endorse 
the viewpoint that this type of critical phenomenon, where  interfaces
 between the two symmetric absorbing states branch and annihilate, is
best 
 described as the {\it (generalized) voter class} 
in the sense of \cite{GV}. Recall that in the usual voter model
\cite{Oldvoter},
  randomly chosen Ising spins take the value of one of their
randomly chosen neighbors: then only  interfaces ($+-$ pairs)
evolve, with $+-\to ++$ or $--$ with equal
probability $\frac{1}{2}$. In dimension $d=2$, this model is critical
and at its upper critical dimension. It is characterized by a marginal
ordering process during which the density of interfaces ($+-$ pairs)
decays like $1/\ln t$ \cite{Oldvoter}.
 In contrast, this simple rule is not critical
in other dimensions: for $d=1$, it coincides with the annihilation
process $2A\to\emptyset$, while in $d=3$ it leads to a disordered
phase.  One possible generalisation of this ``classical'' voter rule
preserving the $Z_2$ symmetry is to allow for spin swaps $+-\to -+$
[which in $d=1$ amounts to the branching reaction $A\to3A$, see
Fig.~\ref{fig1}(a)].  It is simple to realize that this generalisation
(and others) allows for tuning the model at criticality in any
dimension, that in $d=2$ the critical properties of the original model
are preserved, and that in $d=1$ the critical point is nothing but of
the DP2/directed Ising/parity-conserving class. It is thus both
meaningful and useful to denote this class, in any dimension, as the
generalized voter (GV) class.

Extrapolating from their numerical results in $d=2$, Dornic {\it et
al.} \cite{GV} conjectured that transitions with $Z_2$ symmetry and no bulk
fluctuations (i.e.  with two symmetric absorbing states) should all
display GV critical points. In a recent paper, though, Droz, Ferreira,
and Lipowski \cite{drofeli} somewhat challenged this picture
by showing  that such a transition, in some
versions of two-dimensional generalized voter models, may not be direct,
but split into a first Ising-class, symmetry-breaking transition
followed later by a DP-class transition to the absorbing state 
chosen after the previous Ising critical point.  One can thus legitimately
wonder whether the direct GV class transitions observed by Dornic {\it
et al.} ---as well as Droz {\it et al.}--- 
exist at all (at codimension-1 manifolds of
parameter space) or whether they are just the artifact of 
close-by Ising and DP transitions, coinciding only at special points
like the classical voter model.

To summarize, in two space dimensions, the question of the possibility
of the merging of an Ising and a DP line into a full GV line remains open,
 while in one space-dimension the relevance of parity
conservation is still debated. Below, we address both of these points
and clarify the nature of {\it all} phase transitions in the
presence of two symmetric absorbing states via the introduction of a
unique, well-behaved, Langevin equation for this general problem.

Our proposal is by no means unique, but it is constrained by general
guidelines: The equation has to be symmetric under reversal of the
field ($\phi\to -\phi$), which takes values between two absorbing
barriers, set, without loss of generality, at $\pm 1$ ($\phi\in [-1,1]$).
Because, in two dimensions, the transition can be split into an Ising
and a DP point, each of the absorbing barriers must be similar to
those of the Langevin equation for DP, i.e. the square root of the
distance to each barrier must appear as a multiplicative factor of the
noise. This is also corroborated by the fact that the Langevin
equation proposed once for the classical (integrable) voter model \cite{ron}:
\begin{equation}
\label{voter-sde}
\partial_t \phi = D \nabla^2\phi + \sigma\,\sqrt{1-\phi^2}\,\eta
\end{equation}
where $\eta$ is a Gaussian noise delta-correlated in space and time,
 was recently shown to behave as expected ({\it i.e.} logarithmic decay
of the density of interfaces) \cite{HKJ03,Us1}.
In order to represent the possibility of Ising-like spontaneous
symmetry breaking, we need to add a minimal number of polynomial terms
with odd powers of $\phi$.  (At least two free parameters are needed
to describe for the splitting scenario uncovered by Droz {\it et al.}). 
We are then almost ineluctably led to the following equation:
\begin{equation}
\partial_t\phi=(a \phi - b \phi^3)(1- \phi^2)+ D\nabla^2\phi +
 \sigma\,\sqrt{1-\phi^2}\,\eta
\label{Langevin}
\end{equation} 
Note that removing the $1-\phi^2$ factors, both in the deterministic
force and in the noise amplitude, leads to the Model A for the Ising
class \cite{HH}. Let us now describe the different possible regimes of
Eq.~(\ref{Langevin}) in the $(a,b)$ parameter plane, a natural choice
since for $a=b=0$ one recovers the voter equation (\ref{voter-sde}).

We start with a discussion at the mean-field level, i.e. reducing
Eq.~(\ref{Langevin}) to its first term, rewritten as
 $-V'(\phi)$, with the
 ``potential'' $V(\phi)=-\frac{a}{2}\phi^2 +
\frac{a+b}{4}\phi^4 - \frac{b}{6}\phi^6$.

{\it $b >0$: separate Ising and DP transitions.} For $a<0$, $\phi=0$
is locally stable, while it is unstable for $a > 0$
(Fig.~\ref{fig1}b). The $b \phi^3$ term enforces stability as in Model
A (even if the absorbing barriers are removed). At $a=0$, where the
local stability around $\phi=0$ changes, the symmetry is broken, and
we expect an Ising transition in the full problem \cite{GJH}. Increasing $a>0$,
the minima of the potential move progressively closer to the absorbing
barriers and, for $a=b$, a collapse onto the absorbing barrier
selected by the previous spontaneous symmetry breaking takes place. 
This second transition should be in the DP class
once fluctuations are incorporated.

{\it $b\le0$: unique GV transition.}  If $b=0$ the potential is
Gaussian around the origin, which is a stable extremum if $a < 0$, and
unstable otherwise. The transition is at $a=0$, but there is no
$\phi^4$ term in the potential forcing it to be continuous: the
location of the potential minimum changes abruptly from $\phi =0$ to
$\phi = \pm 1$.  This time the symmetry breaking occurs simultaneously
with a fall into one of the absorbing states. The critical point should be
 that of the voter model once fluctuations are
taken into account. For $b <0$, the $\phi^4$ term is
present in the potential, but it is not stabilizing, and it does not
lead to a continuous transition. For $a<0$ the origin is locally
stable, and there are also extrema at the barrier, and additional
maxima at $\pm \sqrt{a/b}$ that may or may not lie in the interval
$[-1,1]$. As $a$ approaches $0$ the extrema move closer to the origin,
and the minima at the barriers deepen [Fig.~\ref{fig1}(c)]. At some
point, the stability is globally changed and we expect to have a
situation similar to the one for $b=0$, i.e. a unique GV transition.

The naive mean-field parameter diagram of Eq.~(\ref{Langevin}) thus
consists of a symmetry-breaking line along the $b$-axis, 
joined by a DP line $a=b$ at the origin. For $b\le0$, the two lines
merge, and a unique GV transition occurs, while for $b>0$ the
Ising-like symmetry breaking does not lead directly to one of the
absorbing states, a situation similar to that uncovered by Droz {\it
et al.}. More elaborated, self-consistent mean-field approaches lead
to similar results, albeit with the symmetry-breaking line not being
along the $b$-axis anymore.

In order to go beyond mean-field and elucidate the influence of
fluctuations in the phenomenology of equation (\ref{Langevin}), we
integrate it numerically using the approach detailed in
\cite{PL,Us1}. This method is designed to circumvent the numerical
difficulties associated with the presence a singular square-root
noise near an absorbing state in Langevin equations. 
It consists in separating the integration of the deterministic terms
from the stochastic piece, the latter
being performed by sampling exactly the conditional probability
distribution function (p.d.f.)  solution of the associated (forward)
Fokker-Planck equation. This sampled value is next used  to
evolve the remaining  deterministic part. 
Here, 
the Fokker-Planck equation for
 $\frac{{\rm d}\phi}{{\rm d}t}=\sigma\sqrt{1-\phi^2}\eta(t)$
can be solved through an eigenfunction expansion, leading to a
rather complicated   p.d.f.,
with  a continuous part and two delta peaks at the barriers
$\phi=\pm 1$ \cite{PL}.  
However, 
as remarked in \cite{Us1}, discarding
the (exponentially suppressed) influence of the farthest absorbing state,
one can treat the noise term as two independent DP barriers, 
  and apply the  existing efficient procedure   for the DP noise. 
 Thereby, using 
time-mesh $\Delta t= 0.1 $, space-mesh $\Delta x= 1$, and the
parameter values $D=0.5$ and $\sigma^2 =0.8$, Eq.(\ref{Langevin}) can
be faithfully integrated.

\begin{figure}
\includegraphics[width=8.6cm,clip]{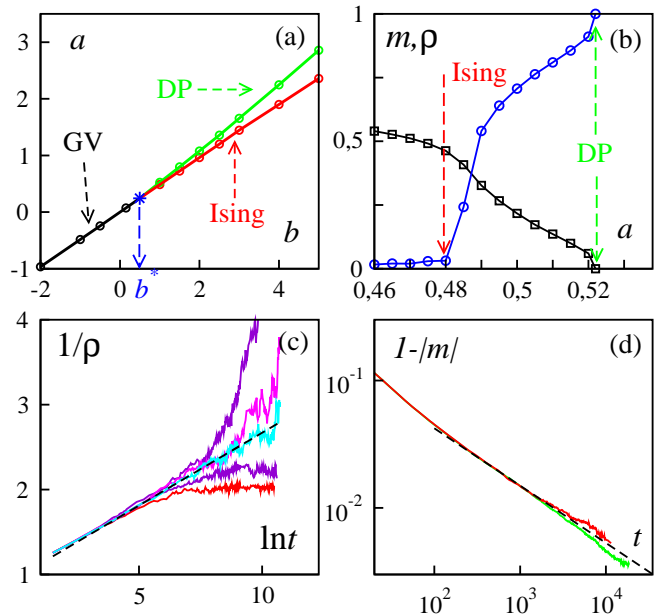}
\caption{(Color online) Results from simulations in $d=2$.
(a) Phase diagram in the $(a,b)$ plane (error bars are smaller than
the symbols size).
  (b)
Steady-state magnetization (circles) and density of interfaces 
(squares) vs $a$ for
$b=1$.  
(c) $1/\rho$ vs $\ln t$ at $b=-0.2$ for various values of $a$ around
$a_{\rm GV}\simeq -0.115$ (middle curve); the dashed line is a linar fit. 
(d) Time-decay of $1-|m|$ for $a$ values around
$a_{\rm DP}\simeq 1.6551$ ($b=3$); at criticality, $1-|m|\sim t^{\theta}$ with
$\theta\simeq\theta_{\rm DP}\simeq 0.45$ (dashed line).  
}
\label{fig2}
\end{figure}

We first present our results obtained in two dimensions, which agree
qualitatively with the phase diagram predicted from mean-field
arguments:

For $b$ larger than some $b^*$ ($b^* \approx 0.50$ with our choice of
parameters), two distinct transitions are encountered upon increasing
$a$ and they merge linearly as $b\to b^*$ 
[Fig.~\ref{fig2}(a) and \ref{fig2}(b)]. 
At low $a$, any initial condition leads to a disordered state
($\langle\phi\rangle=0$). For $a_{\rm Ising}<a<a_{\rm DP}$, the steady
state, reached after some phase ordering transient, has a non-zero
magnetization $m\equiv \langle\phi({\bf r},t)\rangle_{\bf r}$ but is
still fluctuating ($0<|m|<1$) and the density of interfaces
 $\rho=1-\langle\phi({\bf
r},t)\phi({\bf r}+{\bf e}_v,t)\rangle$ (where ${\bf e}_v$ represents any of 
the unit vectors of the underlying square lattice) is finite.
For $a>a_{\rm DP}$,
ordering is complete ($m=\pm 1$, $\rho=0$). We have checked that the
symmetry-breaking transition occurring at $a_{\rm Ising}$ is in the
Ising universality class, both by steady-state finite-size scaling
analysis, and by measuring the decay of the time auto-correlation
function from disordered initial conditions. For instance, the curves for the so-called Binder cumulant 
at different system sizes all cross each other
around the universal value $U^*\simeq0.61$ (not shown).  We have also
checked that the fall into one of the absorbing states is a DP-class
phase transition, as e.g. 
testified by the algebraic decay in time of the activity
in a large system after a critical quench [Fig.~\ref{fig2}(d)].

For $b<b^*$, a unique transition is observed at $a=a_{\rm GV}$, across
which the steady-state magnetization jumps from zero to $\pm 1$. On
the other hand, the density of interfaces goes continuously
to zero as $a\to a_{\rm GV}$ from below (not shown). 
 At $a=a_{\rm GV}$, the logarithmic time decay of $\rho$ is one of the
 hallmarks of the GV  class [Fig.~\ref{fig2}(c)].

\begin{figure}
\includegraphics[width=3.8cm,clip]{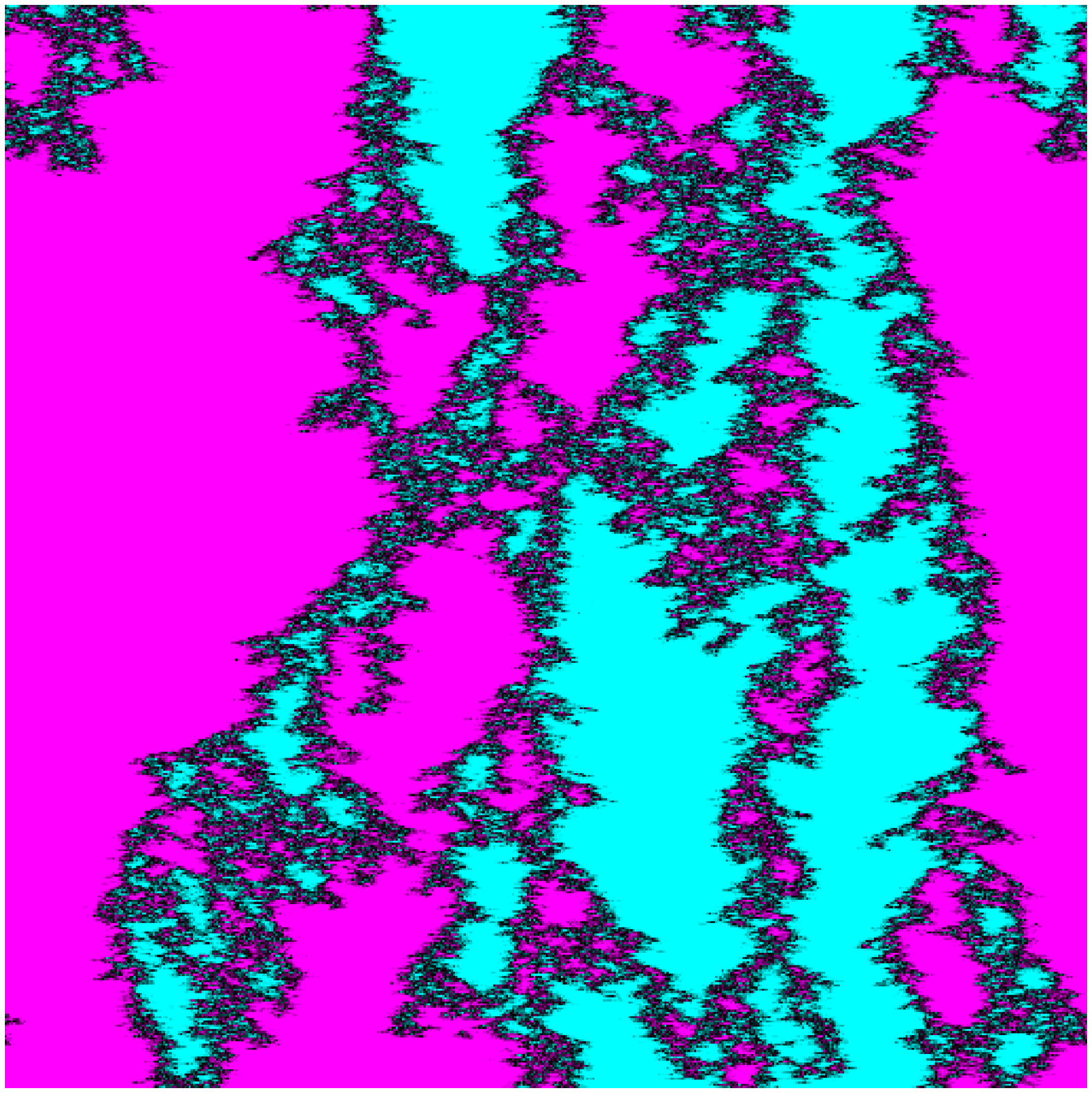}
\includegraphics[width=4.2cm,clip]{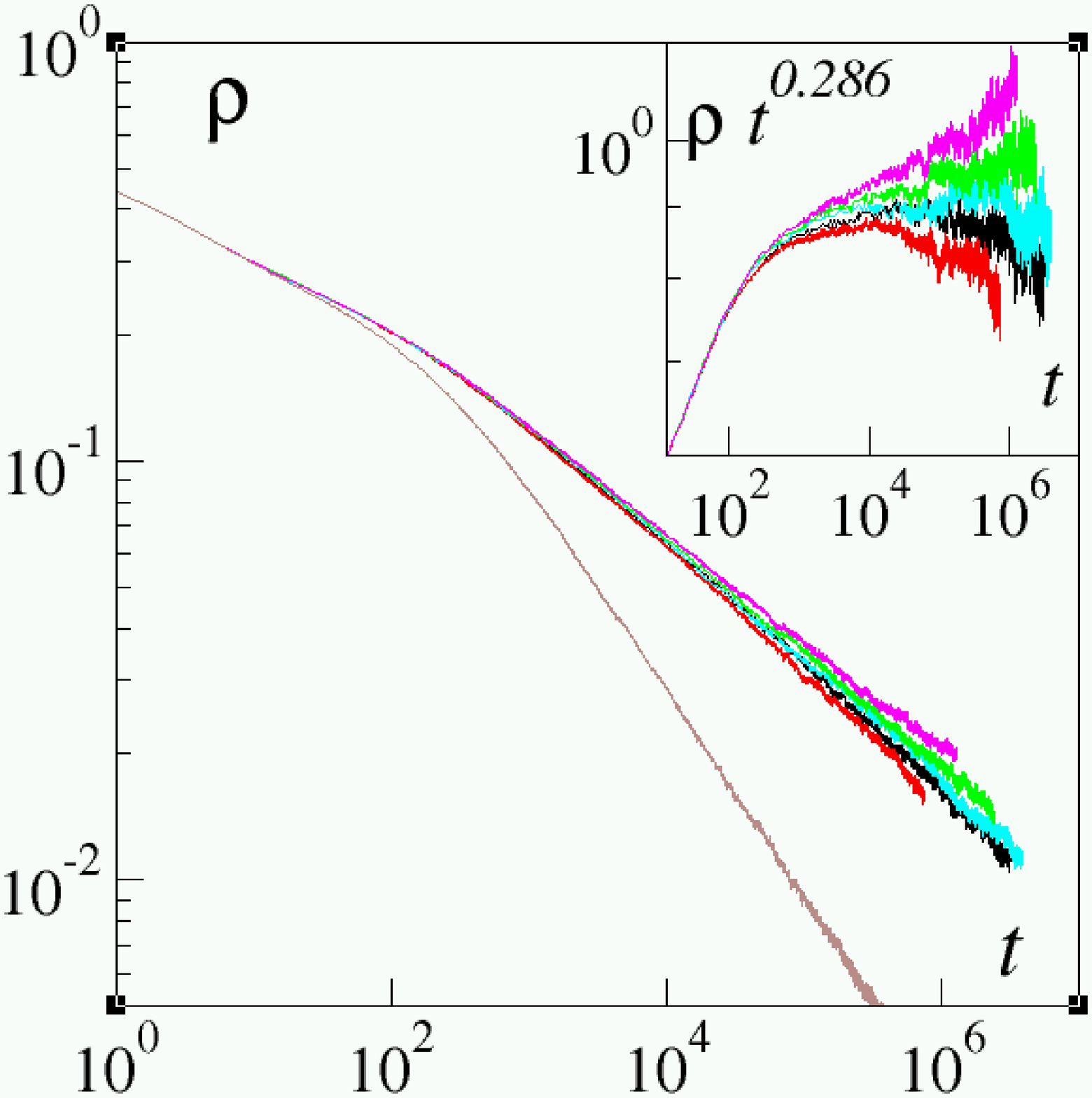}
\caption{(Color online) Results from simulations in $d=1$.
(a) space-time plot of $\phi$ at criticality $a=-0.1255, b=-0.25$, 
system size 512, time running downward for $10^4$.
(b) time decay of the interface density  
$\langle\rho\rangle$ for $b=-0.25$ and, from bottom up:
 $a=-0.12$, -0.124, -0.125, $a_{\rm GV}\simeq -0.1255$, -0.126, -0.127;
in the absorbing phase ($a=-0.12$), $\rho\sim 1/\sqrt{t}$;
inset: data around   $a_{\rm GV}$ multiplied by $t^{0.286}$ 
(system size $2^{21}$, $\Delta x=1$, $\Delta t=0.1$, $D=\sigma=0.25$, 
initially, $\phi=0$ everywhere).
}
\label{fig3}
\end{figure}

In one space dimension, where general arguments exclude the existence
of an Ising transition, we expect a unique, {\it continuous}
direct transition from a
disordered phase ($m=0, \rho>0$) to one of the two absorbing states
($m=\pm 1, \rho=0$). It is therefore near at hand to surmise that
the ensuing critical point should be characterized by the
well-known \cite{Review_Haye,Review_Odor} 
critical exponents of the GV class. A direct integration of
Eq.(\ref{Langevin}) fulfills all these expectations: a space-time plot
of $\phi$ reveals the typical branching-annihilating dynamics of kinks
(Fig.~\ref{fig3}a) schematized in Fig.~\ref{fig1}a.  At criticality,
the density of interfaces decays with an exponent $\theta =0.28(1)$
fully compatible with the expected value $0.286$ (Fig.~\ref{fig3}b).

In three dimensions, preliminary results indicate that the scenario
observed in $d=2$ remains valid: for large $b >0$, separate
 Ising and  DP
transitions are found, while for (essentially) negative $b$ values,
a unique, direct transition occurs. This last transition, however,
does {\it not} possess the marginal character of its two-dimensional
counterpart: indeed it is ``fully first-order'', in the sense that
the density of interfaces now also jumps discontinuously to zero. We note
that this is in agreement with $d=2$ being the upper critical dimension
for the GV class.

We now summarize and comment on our results. 
The Langevin equation
(\ref{Langevin}) does account for all the known phenomenology about
order-disorder phase transitions in the presence of two symmetric
absorbing states. In $d=1$, such a description puts an end to the
discussion, in microscopic models, about the relevance of ``parity
conservation''. We note {\it en passant} that in the absorbing phase of the
GV transition our equation becomes an effective continuous description
of the annihilation process $2A\to\emptyset$, as testified by the
$1/\sqrt{t}$ decay in Fig.~\ref{fig3}b.  In $d=2$, we obtained clear
evidence that a full GV line exists, but splits into an Ising and a DP
line at some parameter value $b^*$ (inset of Fig.~\ref{fig2}a), in
qualitative agreement with mean-field predictions.  In $d=3$, 
and more generally above the upper critical dimension of the GV class,
the same scenario is found but for the fact that the GV transition is
now fully first-order.

It is not clear to us whether  Eq.~(\ref{Langevin}) can be derived
from first-principles, especially given the importance,
in microscopic models exhibiting the GV transition,
 of the annihilation reaction $2A \to \emptyset$,
 for which  this  is reputedly impossible  \cite{NoLangevin}. 
Nevertheless,   both the generating functional associated to Eq.~(\ref{Langevin})
and the one rigorously obtainable from the master equation
for the corresponding reaction-diffusion processes
 enjoy the same symmetry and feature similar characteristic invariants.
Thus analytical studies of our proposal appear promising, 
be it either within renormalisation-group perturbative calculations (maybe
akin to those of \cite{Cardy00}),
 or  via the nonperturbative approach  put forward in \cite{Canet}.
First results under the latter auspices are very encouraging \cite{TBP}.

We  warmly thank P. Grassberger for  stimulating
 discussions.
M.~A.~M. acknowledges financial support from the Spanish MCyT (FEDER)
under project BFM2001-2841. I.D. is  also indebted to the Service de Physique
Th\'eorique (CEA Saclay) for  generous financial support.

\end{document}